\documentclass[10pt,a4paper]{article}
\usepackage{makeidx,latexsym}
\usepackage{color,graphicx}
\usepackage{makeidx}
\title{Recent updates on the ArDM project: A Liquid Argon TPC for Dark Matter Detection}

\author{V. Boccone\footnote{Physik-Institut der Universit\"at Z\"urich,  Winterthurerstrasse 190, CH-8057 Z\"urich, Switzerland. E-Mail:boccone@cern.ch}, on behalf of the ArDM collaboration. }
\date{June 15$^{th}$ 2008} 
\begin{document}
\maketitle
\begin{abstract}
ArDM is a new-generation WIMP detector which will measure simultaneously light and charge from scintillation and ionization of liquid argon. Our goal is to construct, characterize  and operate a 1 ton liquid argon underground detector. The project relies on the possibility to extract the electrons produced by ionization from the liquid into the gas phase of the detector, to amplify and read out with Large Electron Multipliers detectors. Argon VUV scintillation light has to be converted with wavelength shifters such as TetraPhenyl Butadiene in order to be detected by photomultipliers with bialkali photocathodes. We describe the status of the LEM based charge readout and light readout system R\&D and the first light readout tests with warm and cold argon gas in the full size detector.
\end{abstract}

\section{Introduction}
Noble liquids such as argon or xenon are today two of the best options for large-size Dark Matter (DM) experiment such as WIMPs (Weak Interacting Massive Particles) searches because they serve as target and detecting media at the same time. WIMPs in fact, with particular regard to the lightest neutralino, are the most popular DM candidates within the Minimal Supersymmetric Standard Model (MSSM). Most of the DM models predicts that WIMPs interact almost only gravitationally and maybe weakly with matter thus forming cold relic halos around massive astronomical objects. For this reason DM is very difficult to detect directly nevertheless its evidence has been observed using astronomical observation \cite{xchandra}.
 
Many groups in the world are focusing their attention on the direct detection. The detectors should have big target masses and excellent noise rejection capabilities because of the small cross section between DM and ordinary matter. Liquid noble gases (Ar, Xe, Kr, Ne) have a relatively low ionization energy and a good cross section for nuclear recoil processes, moreover they can act as active target because they not only have good scintillation properties but also a long electron lifetime. Argon, because of its availability and low cost, is the best one among those candidates and is very competitive for large detectors. 

\subsection{Interaction with liquid argon}
\begin{figure}[h]\begin{center}
\includegraphics[height=4cm]{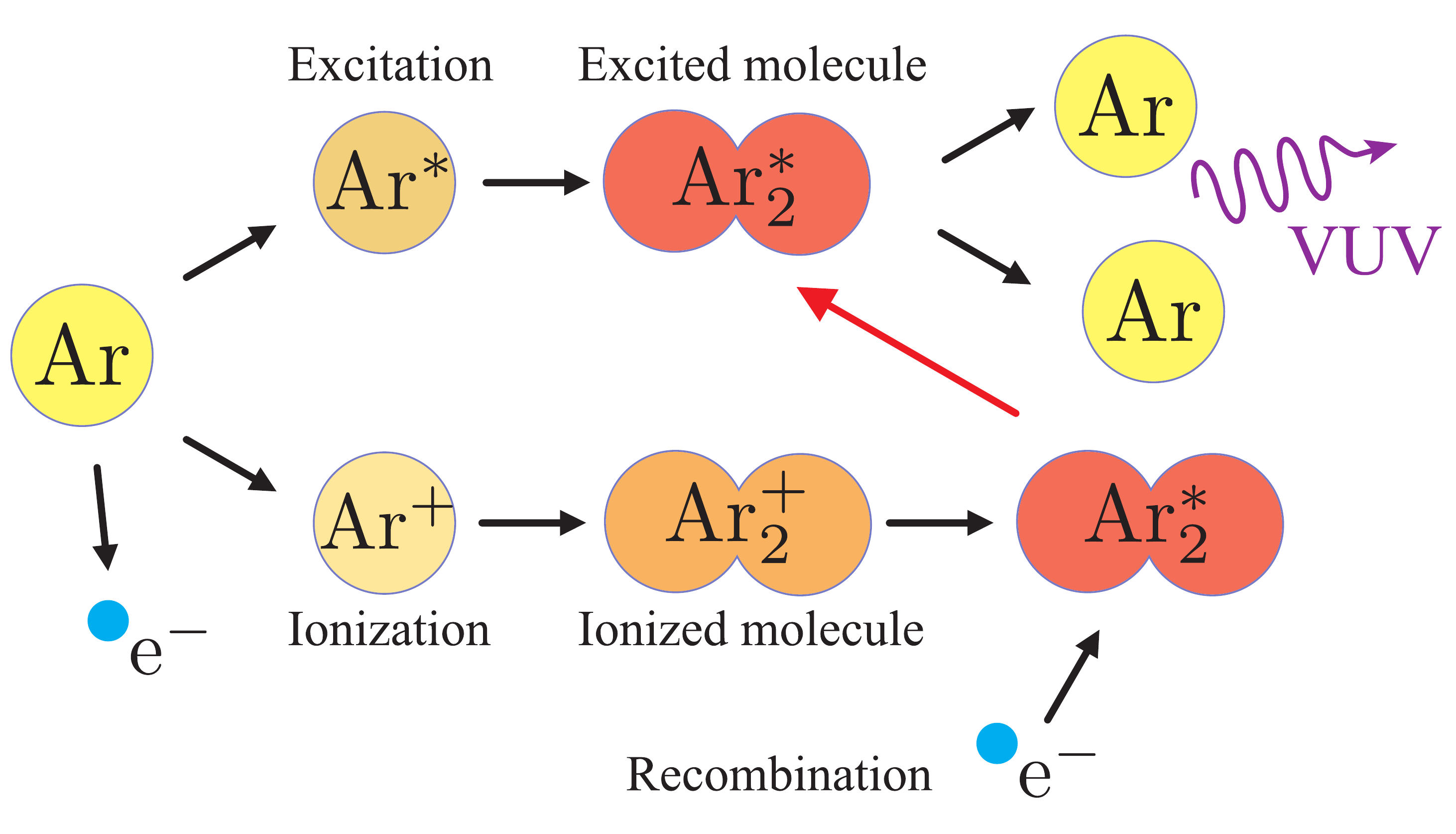}
\caption{\label{argonscintillation} Scintillation and ionization in argon.}
\end{center}\end{figure}

The expected DM signal from a WIMP-argon head on collision is a single 10-100keV recoiling nucleus. The passage of ionizing particles in argon produces not only ionization but excitation of atoms as well. Ionized and excited atoms form the excited molecular states Ar$^{+}_{2}$ and Ar$^{*}_{2}$ respectively.  Ar$^{+}_{2}$ eventually recombines with an electron, producing Ar$^{*}_{2}$ in the molecular excited state which decays radiatively (see fig.\ref{argonscintillation}).

The lowest allowed radiative decays from the molecular excited states are the transitions from the  singlet and the triplet states ($^{1}\Sigma^{+}_{u}$ and $^{3}\Sigma_{u}^{+}$) to the dissociative ground state ($^{1}\Sigma^{+}_{g}$) which consist of two independent atoms; the emission spectrum for those transitions has a narrow peak at $(128\pm10)$nm in the Vacuum Ultra Violet (VUV) band \cite{larlight1}. Reabsorption cannot occur because the energy of the single atomic excited state is too high. Only impurities such as water, N$_{2}$ and CO$_{2}$ can eventually reabsorb VUV light and capture the electrons.

Those two transition have two different decay times, respectively $\tau_{1}$ and $\tau_{2}$, which do not depend on the type or on the energy of the incident particle; $^{1}\Sigma^{+}_{u}\rightarrow $$^{1}\Sigma_{g}^{+}$ is strongly allowed and has a short decay time while $^{3}\Sigma^{+}_{u}\rightarrow$$^{1}\Sigma_{g}^{+}$ is allowed only because of spin-orbit coupling and has a much longer life. In liquid argon (LAr) $\tau_{1}=5$~ns and $\tau_{2}=1.6$~$\mu$s.

\begin{figure}[h]\begin{center}
\includegraphics[height=3cm]{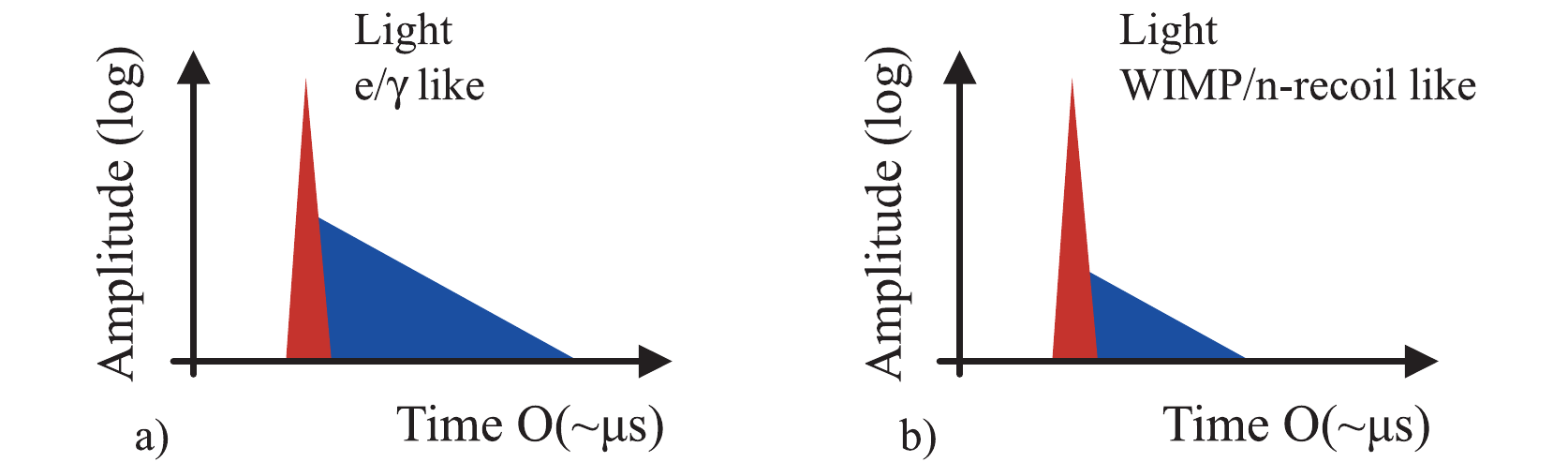}
\caption{\label{argontopologyA} Argon scintillation event topologies.}
\end{center}\end{figure}

 The ratio between the populations of the two states $r_{ex}= \delta( ^{1}\Sigma^{+}_{u}) / \delta( ^{3}\Sigma_{u}^{+})$ depends strongly on the ionization density of the track (see fig.\ref{argontopologyA}); for light projectiles such as electrons and $\gamma$s $r_{ex} \approx 1/2$ while for $\alpha$s and fission fragments is $r_{ex} \approx 4\div5$. The electron lifetime and the decay time of triplet state strongly depend on the on the purity of the liquid.

\section{The ArDM experiment}
\subsection{Detector description}
\begin{figure}[h]\begin{center}
\includegraphics[width=0.8\textwidth]{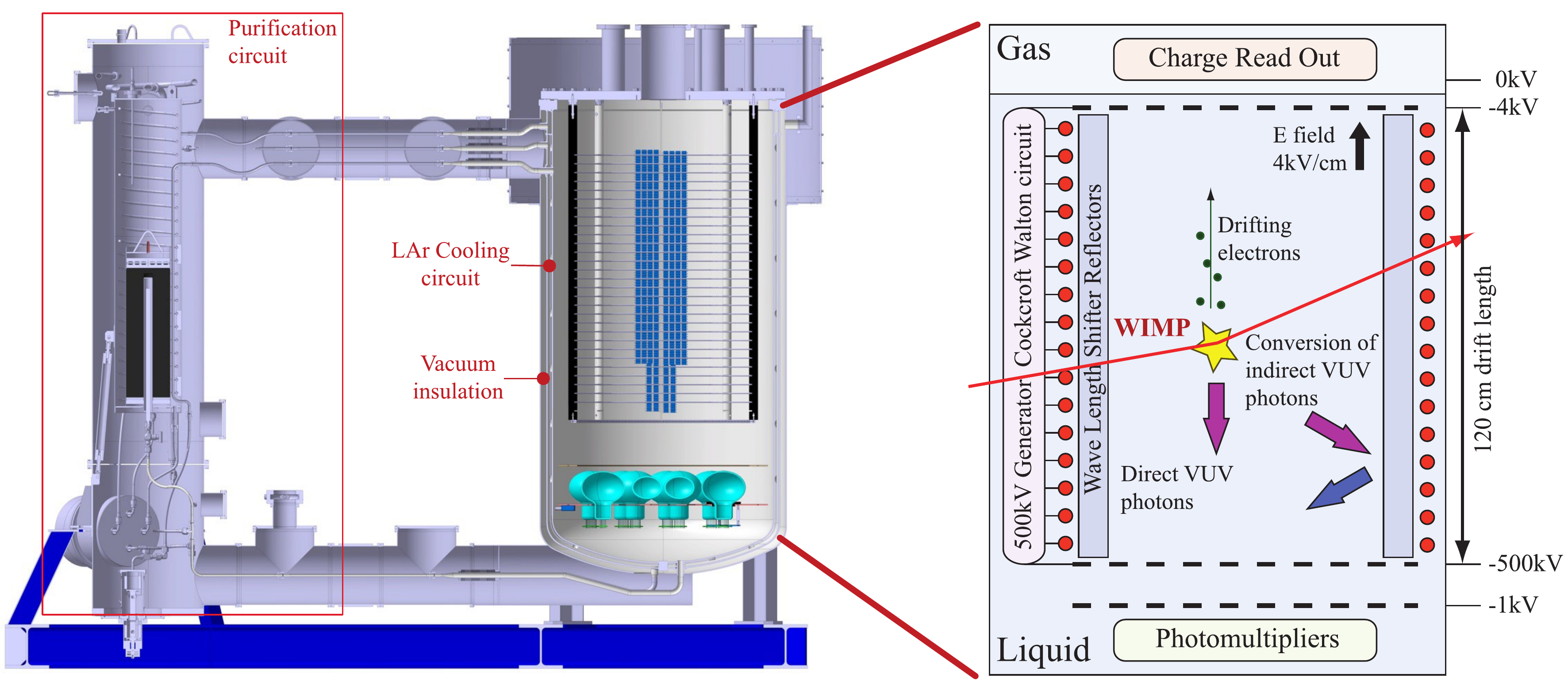}
\caption{\label{ArDM} Conceptual layout of the ArDM experiment.}
\end{center}\end{figure}

The conceptual layout of the ArDM experiment \cite{Rubbia:2005ge} is illustrated in fig.\ref{ArDM}. An ionizing particle traveling through an active mass of $850$~kg generate scintillation and ion pairs; a 500 kV Cockcroft Walton generator chain provides the electric field necessary to drift the electrons towards the gas phase on the top of the detector. The electrons are then extracted from the liquid to the gas phase ($E_{extr}\sim$3 kV/cm) and accelerated towards the double stage Large Electron Multiplier (LEM) detectors which provides multiplication (gain $\approx 10^4$) and position reconstruction. 
Detection of light is performed by an array of 14 cryogenic photomultipliers (PMTs) located below the cathode. Most of the VUV scintillation light hits the side of the detector, therefore we decided to install wavelength shifter (WLS) coated diffusive reflectors which shift the light from 128 nm to 420 nm ($\lambda_{peak}$ for bialkali photocathode) and diffuse the blue light in the detector. A WLS coating of the PMT surface is also foreseen to detect the direct light which is $\sim$10\% of the total light. We plan to achieve an average light collection efficiency of 2\%. With a combined light/charge analysis we should reach an energy threshold of approximately \mbox{30 keV}.

In order to reach this goal it is necessary to have a very high LAr purity and an excellent background rejection level. A good purity is obtained by recirculating the LAr through an active copper filter (purification  circuit) with a specially developed cryogenic pump. Argon quality is monitored using two independent purity monitoring systems, which measure the electron drift length and the decay time of slow component ($\tau_{2}$). The main vessel and the purification circuit are surrounded by an insulation vacuum; a LAr circuit embedded in the walls of the inner vessel allows to pre-cool the detector.

\subsection{Background sources} 
Common background sources for this type of detectors are neutrons, electrons and $\gamma$s.
U and Th chains contaminating the detector components produce neutron and $\gamma$ background, P isotopes (present in steel) have $\beta$ activities. Muons induce neutrons and, if interacting with argon, produce the unstable long living $^{39}$Ar isotope. LAr obtained from the liquefaction of air contains this isotope in a non negligible fraction; the WARP collaboration has reported an activity of \mbox{$1$ Bq/kg} \cite{argon39} which then makes $^{39}$Ar one of the most problematic source of background for ArDM.

The rejection of this background is a complicated issue and involves more detailed studies which were the subject of simulations.  

\begin{figure}[t!]\begin{center}
\includegraphics[height=3cm]{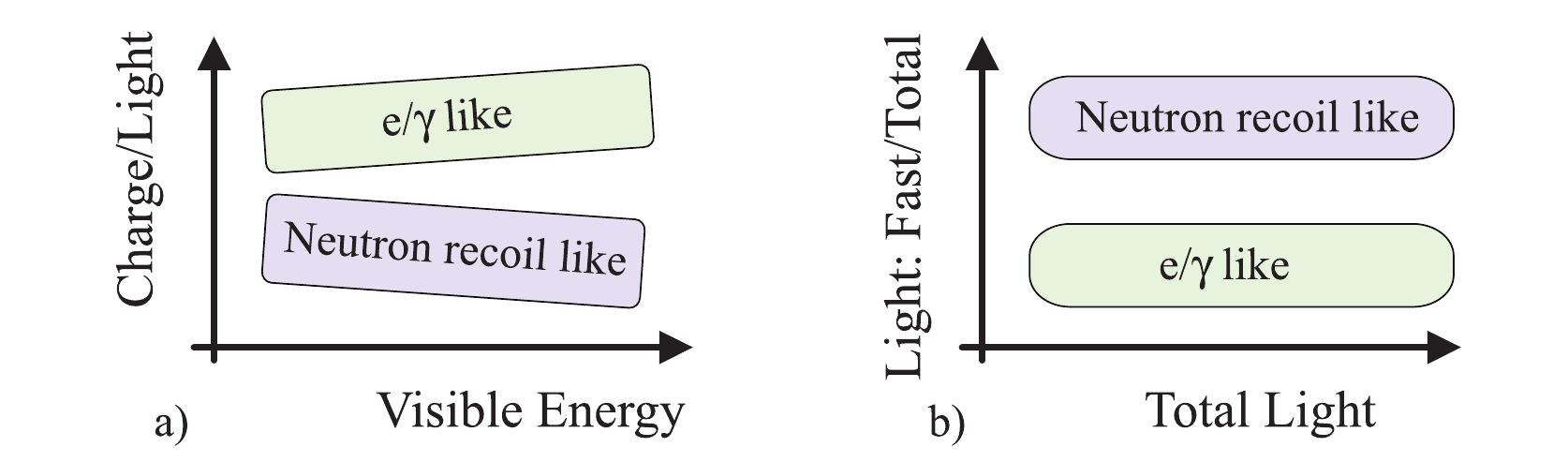}
\caption{\label{argontopologyB} Event topologies.}
\end{center}\end{figure}

Since we measure the charge with its position, time distribution and the scintillation light with its time structure down to {\it ns} scale we are able to distinguish between electrons, $\gamma$s and recoiling nuclei.
An example of rejection studies can be seen in fig.\ref{argontopologyB}: it is possible to identify electron and $\gamma$ events by looking at the ``Charge/Light ratio'' as a function of the total reconstructed energy (fig.\ref{argontopologyB}a) or if we want to use only the light information (for instance to reduce the trigger rate) we can measure the fraction of light coming from the fast component as a function of the total light (fig.\ref{argontopologyB}b). This kind of analysis is fundamental to reject $\beta$ radioactivity from $^{39}$Ar.
 
WIMPs and neutrons lead to similar event topologies since they produce a recoiling argon nucleus; however the energy spectra and the scattering multiplicities are different. We can partly reject neutrons because more than 50\% of them scatter more than once; moreover only 10\% of them produce a single recoil event in the energy range between \mbox{30 keV} and \mbox{100 keV}.

We estimate with a MonteCarlo the total number of neutrons per year coming from the detector components; from simulation we expect $\sim$1000 neutrons/year which will generate a WIMP-like recoil rate of $\sim$50 fake WIMPs/year.
 
\subsection{Charge readout system}
The LEM is a scaled-up version of the Gaseous Electron Multiplier (GEM) (\cite{sauli}, \cite{peskov} and \cite{chechik}).\newline
We are developing a double stage LEM system composed of two LEM planes and a segmented anode plane; each LEM plane consists of a dual layer \mbox{$1.5$ mm} thick printed circuit board (PCB) with \mbox{$0.5$ mm} diameter holes with a pitch of \mbox{$0.8$ mm}, made by precise machining (standard PCB technology). Holes in the last plane are grouped by strips similarly to the anode plane in order to provide a 2D segmented readout of ions and electrons. The third dimension along the cylinder is obtained by time projection imaging. A granularity of few millimeter will grant a sufficient vertex resolution for rejection of multiple scattered neutrons.

\begin{figure}[bt!]\begin{center}
\includegraphics[width=0.8\textwidth]{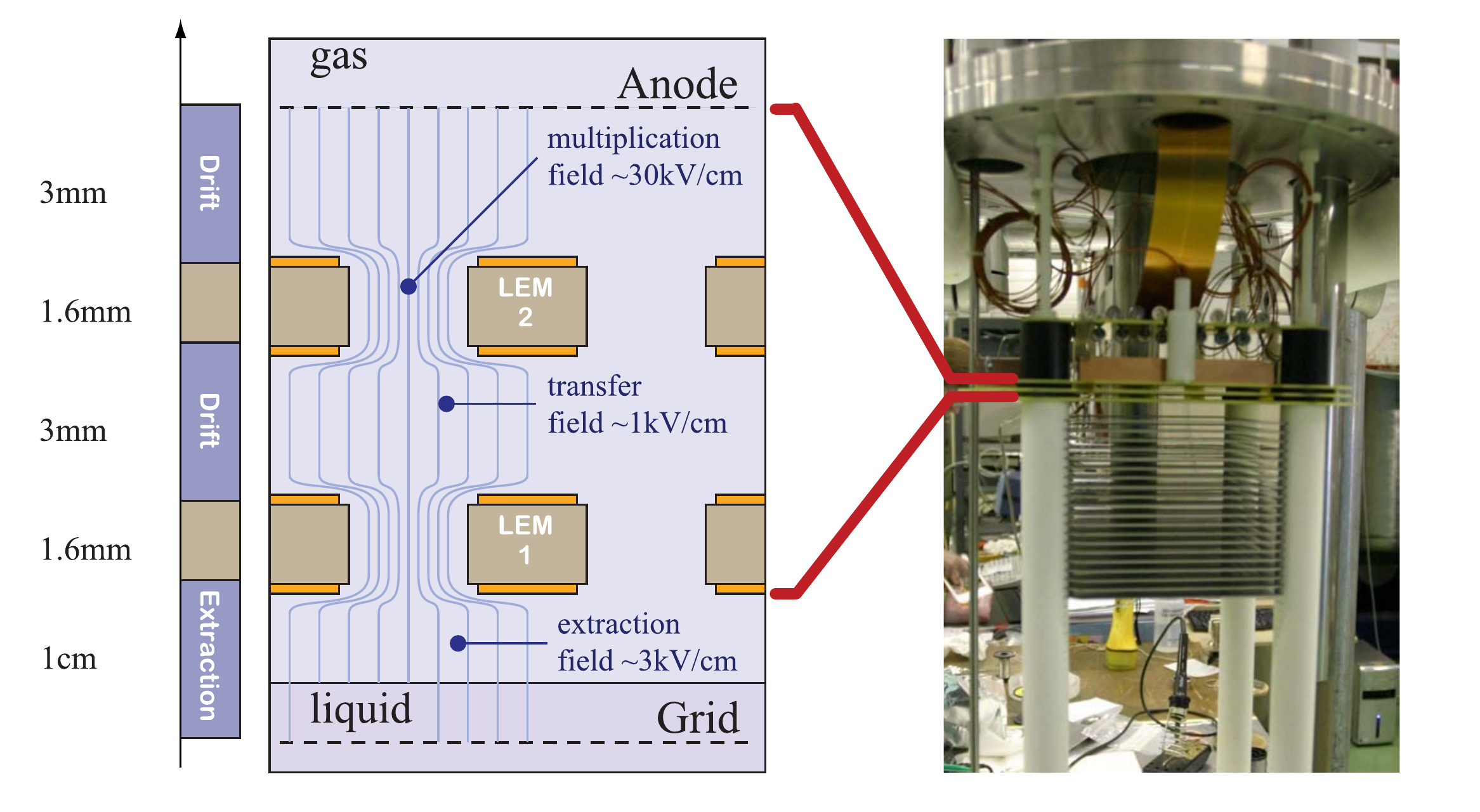}
\caption{\label{LEMp}{\bf Left}: Schematic of the $10\times 10 ~\rm{cm}^{2}$ dual LEM prototype layout. {\bf Right}: Picture of the small R\&D dual LEM.}
\end{center}\end{figure}

We have already achieved operation of a dual stage LEM system in dual phase argon, extracting electron from the liquid to gaseous phase with a small \mbox{10x10 cm$^{2}$} prototype in a dedicated setup (fig.\ref{LEMp}).

\subsection{Light readout system}
The light read out system must be able to detect and identify signals in a range from the single to few thousand photons. It is composed mainly of two parts, the reflectors and the detectors, which have been subject of dedicated R\&D projects (\cite{amslerargon} and \cite{bueno}). 

The internal surface of the active volume is covered by WLS reflector panels. We selected TetraPhenyl Butadiene\footnote{1,1,4,4-TetraPhenyl-1,3-Butadiene.} (TPB) as the best WLS for VUV to blue light conversion because of its light conversion efficiency and its diffusive reflection properties. We optimized the WLS surface density on different reflector materials in a small 1lt liquid argon setup in gas and liquid. A surface density of \mbox{1 mg/cm$^{2}$} was evaporated on \mbox{250 $\mu$m} thin Tetratex\footnote{High performance ePTFE membrane.} membrane. Each reflector \mbox{($130~\times~30$ cm$^{2}$)} was later mounted on 3M ESR\footnote{3M Vikuiti\texttrademark Enhances Specular Reflection.} foils which provide mechanical support and shielding from the light coming from the nonactive volume. We prepared and installed in the detector 15 WLS reflector panels.

\begin{figure}[tbh!]\begin{center} 
\includegraphics[height=4.5cm]{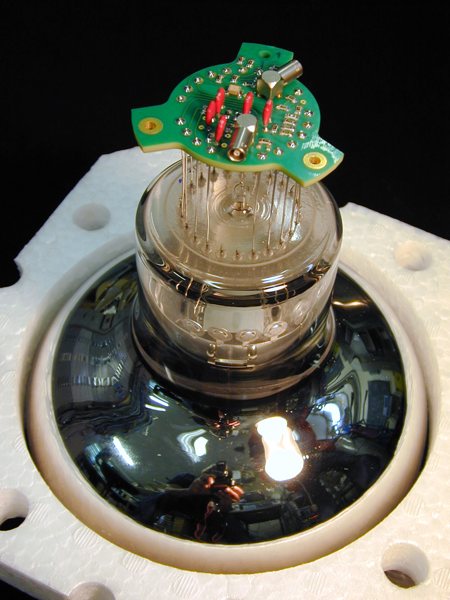}\hspace{0.5cm}
\includegraphics[height=4.5cm]{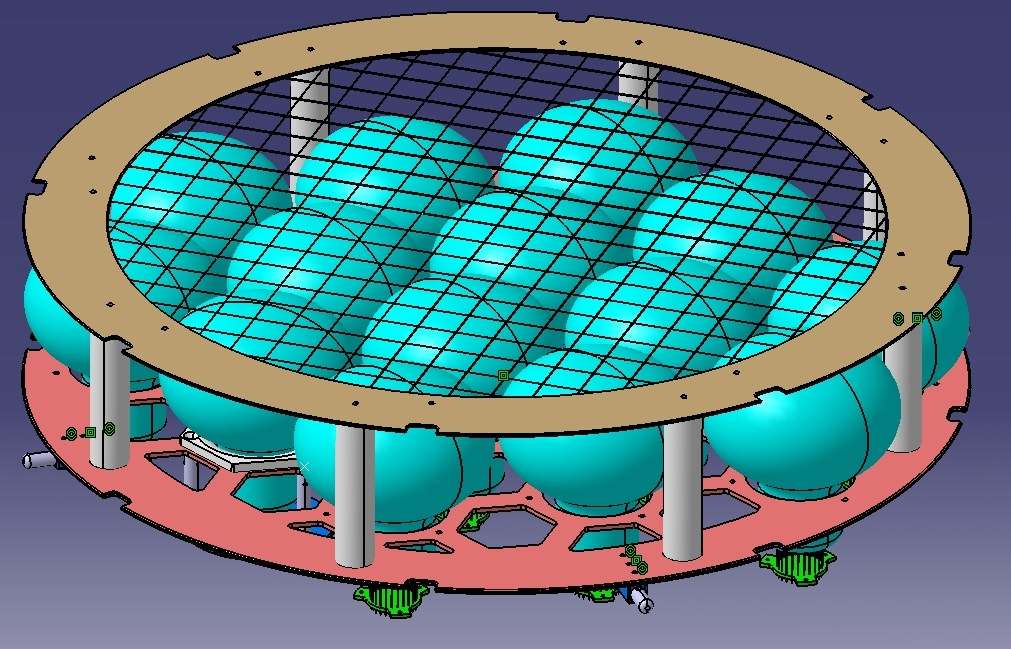}
\caption{\label{pmtmulti} {\bf Left}: Cryogenic PCB mounted on the photomultiplier. {\bf Right}: Design of the light detector.}
\end{center}\end{figure}

The photon detector is composed of 14 special cryogenic PMTs with platinum underlay below the bialkali photocathode, which allows charge restoration at cryogenic temperatures. 
A low activity glass is desired to reduce the amount of self background. Each PMT is mounted on a 3 mm thick cryogenic PCB base which provides voltage supply, signal extraction and mechanical support. A detail of the prototype of the PCB base mounted on the photomultiplier can be seen in fig.\ref{pmtmulti} (left). 

 Polyethylene collars positioned on the steel support frame keep the PMTs in the horizontal position at the right distance from the others. The PMTs are shielded from the \mbox{$-500$ kV} of the cathode with a grid placed at the same potential of the PMT photocathodes. The mechanical structure can be seen in fig.\ref{pmtmulti} (right).
\section{Test of light readout in argon gas}
\begin{figure}[thb!]\begin{center} 
\includegraphics[width=0.8\textwidth]{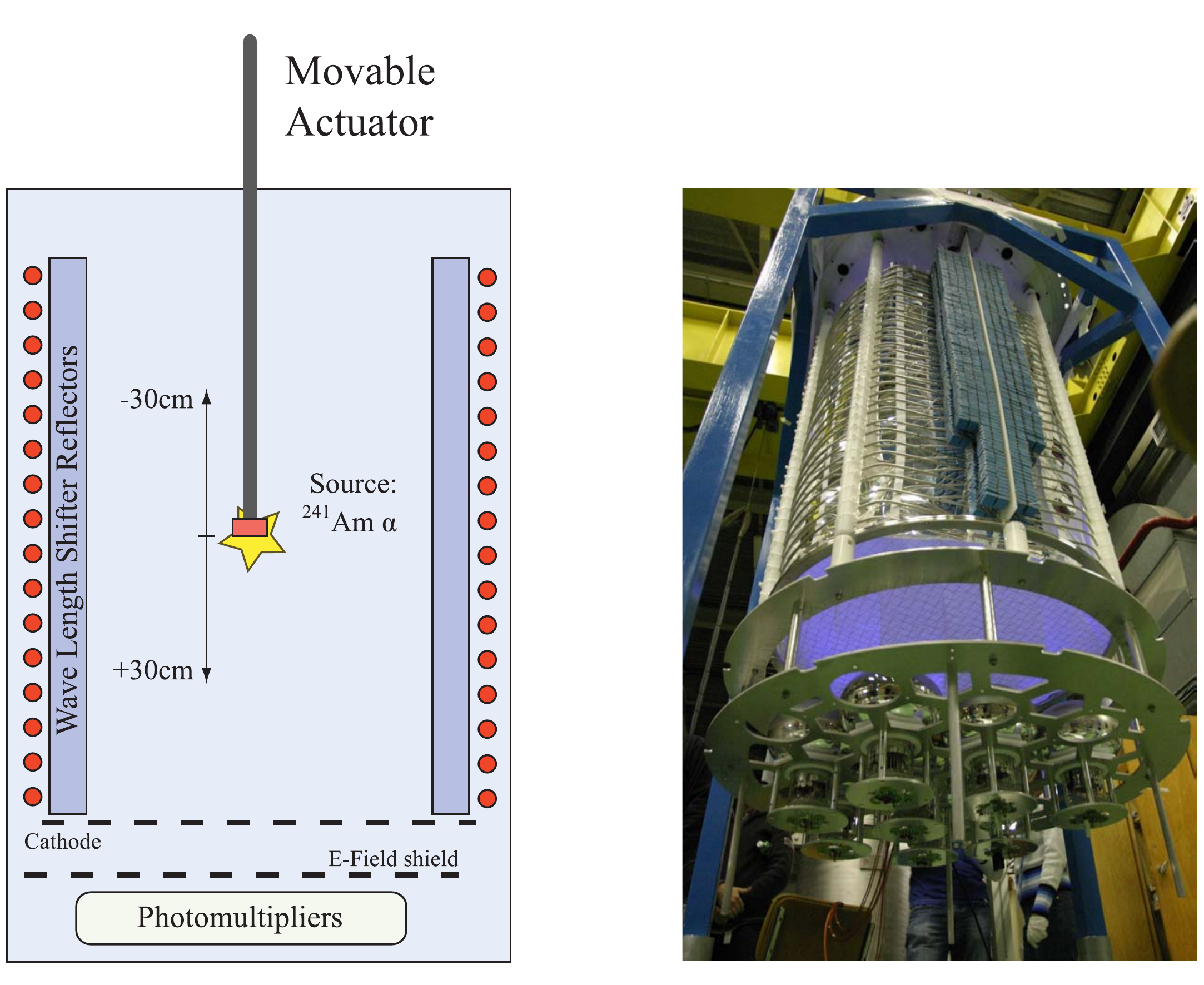}
\caption{\label{setupt} {\bf Left}: Scheme of the setup used in the first tests. {\bf Right}: Photograph of the assembled setup with internal UV light illumination.}
\end{center}\end{figure}

For the test in gas and liquid argon we installed the 15 light shifter reflector panels (full coverage) and 8 of the 14 cryogenic PMT modules.
We have seven Hamamatsu R5912 PMTs (five with 14 dynodes and two with 10 dynodes) and one ETL 9357 KFLB; both types were studied at cryogenic temperatures \cite{raselli} \cite{bueno}. The PMT modules were tested in LAr (88K) with a blue LED \mbox{($\lambda_{pk}=400$ nm)} light before coating and assembly in the experiment. A brief list of the PMT properties is listed in table \ref{tabpmt}. 
\begin{table}[ht]\begin{center} 
\begin{tabular}{|llccc|}	\hline
Producer  & Model       & Quantity & Dynodes & Max. Gain (1500V)\\ \hline
Hamamatsu & R5912/02mod &    x5    &  14      &$\approx 1 \cdot 10^{9}$   \\
Hamamatsu & R5912/01mod &    x2    &  10      &$\approx 1 \cdot 10^{7}$   \\
ETL       & ETL9357     &    x1    &  12      &$\approx 1 \cdot 10^{7}$   \\ \hline
\end{tabular}
\caption{Types of photomultiplier used in the tests.}
\label{tabpmt} 
\end{center}\end{table} 
The E-Field protection grid (which will protect the PMT surfaces from the $-500$ kV potential of the cathode) and the electric field generator chain were installed, but not used for these tests. Ten PT-1000 temperature sensors, placed at different depth in the detector, allowed precise measurements of the inner temperature of the detector. An $^{241}$Am $\alpha$-source (40kbq) and two blue LEDs were installed on a movable magnetic actuator placed on the center of the main flange, since the LEM was not present. The $\alpha$ source has a very thin metallic window which led to a reduction of the $\alpha$ energy from \mbox{5.486 MeV} to \mbox{$\sim$4.5 MeV}.
The test setup is schematized in fig.\ref{setupt} (left) while the picture of the assembled setup illuminated by UV light from inside is shown in fig.\ref{setupt} (right).

\begin{figure}[tb!]\begin{center} 
\includegraphics[width=0.8\textwidth]{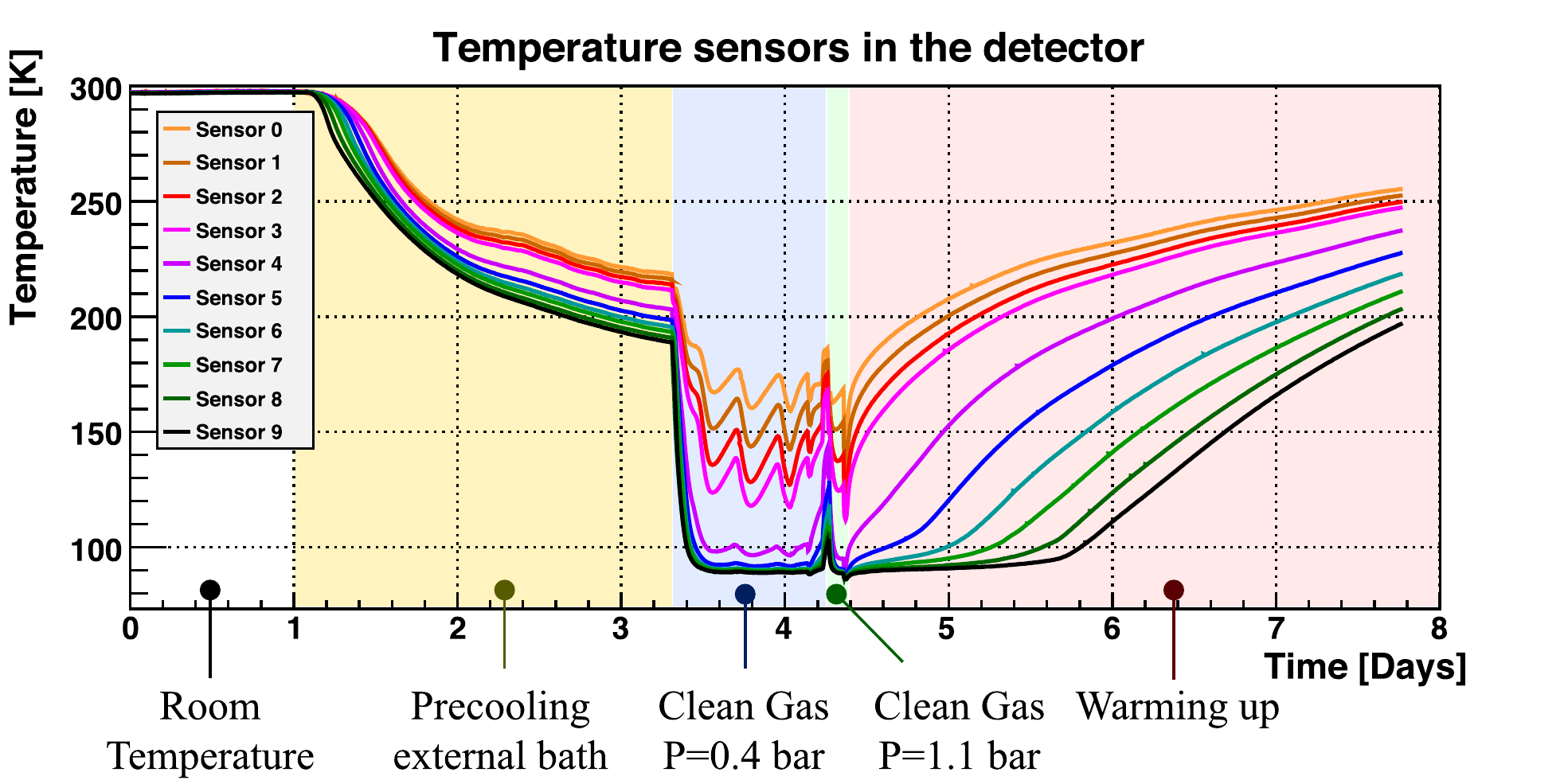}
\caption{\label{cooldown} Time variation of the temperature inside the vessel during the light readout test period. Sensors are distributed along the detector height from top to bottom; sensor 9 is placed at the level of the PMT photocathodes.}
\end{center}\end{figure}

\begin{figure}[thb!]\begin{center} 
\includegraphics[width=1 \textwidth]{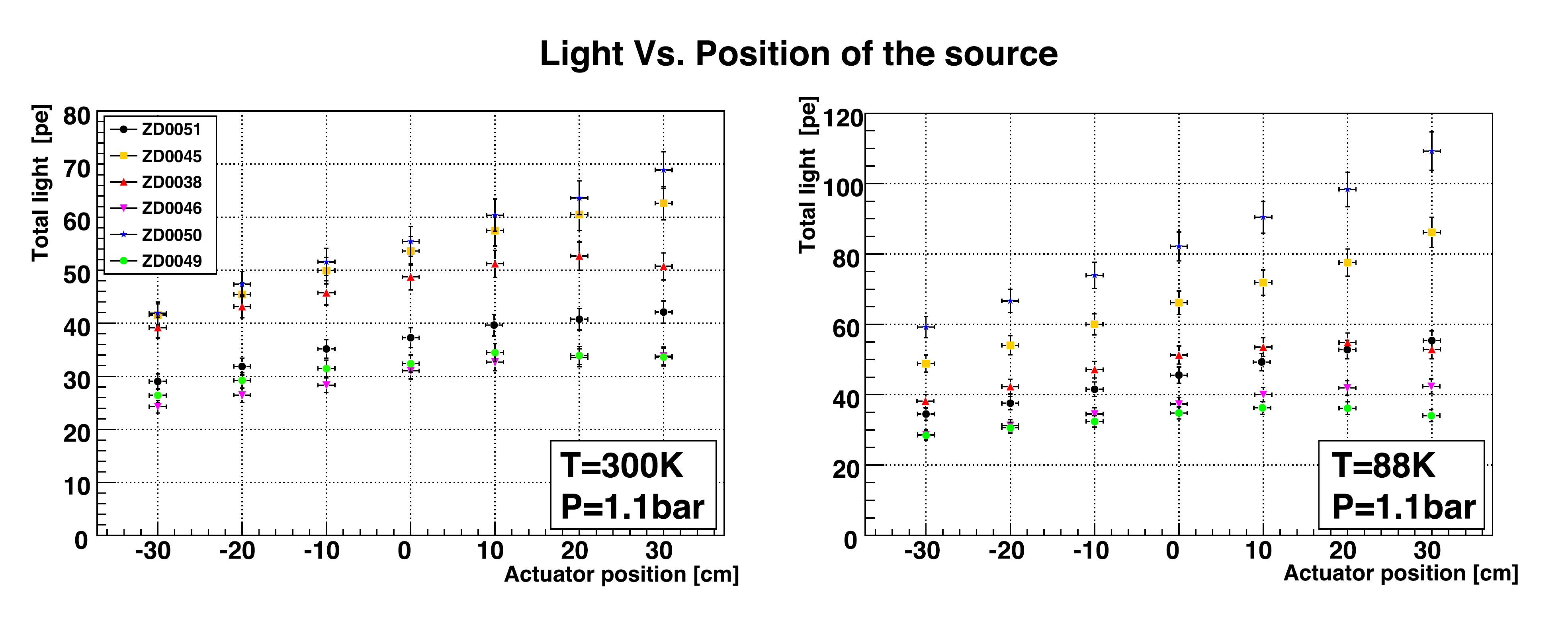}
\caption{\label{ly} Light collected from the PMTs at different source positions in gaseous argon in the warm (left) and cold (right) condition.} 
\end{center}\end{figure}

The PMTs were tested in gaseous argon at room and cryogenic temperatures. The setup was first closed and evacuated down to \mbox{$10^{-6}$ mbar}. We then filled the dewar with gaseous argon\footnote{Ar60, from Carbagas, impurities $\sim$ 1 ppb.} up to a pressure of 1.1 bar. Several measurement of gain, dark count rate and light yield as a function if the source position have  been performed and repeated fig.\ref{ly} (left); two PMTs were not working in gas because of sparking in the high voltage power supply circuit.
Then the setup was evacuated down to $10^{-6}{\rm mbar}$ and liquid argon was circulated in the cooling circuit for more than two days. We filled then the detector with gaseous argon up to a pressure of 0.4 bar, turned on the PMTs and let the temperature sensors inside the dewar stabilize over-night to 88K. Meanwhile a set of data was acquired each hour to study possible effects of gain drift and quantum efficiency (QE) change on the PMTs. The temperature variation as a function of time inside the detector is shown in fig.\ref{cooldown}. 

At the beginning of day 4 temperature was stable at the photocathode level; we repeated the measurement of gain, dark count rate and light yield as a function of source position in cold argon gas at \mbox{P=1.1 bar} fig.\ref{ly} (right). We measured a light yield increase of $\sim$50\% when passing from room temperature to argon gas at \mbox{88 K}. This effect has not been yet fully understood, but is mainly related to the emission spectrum of the gaseous argon at those thermodynamic conditions. A detailed analysis will be performed with LAr for which we will be able to compare directly the light yield with the one we obtained in the small LAr cell.

\section{Conclusion}
In conclusion I report about the progresses done recently by the ArDM collaboration. Many R\&D efforts have been invested in the last 3 years to develop a suitable charge and light readout which fulfills our requirements. ArDM is now in the commissioning phase; the light readout has been installed in its prototype version, the choice of the final PMTs will be done after the liquid argon test. The full size prototype of the LEM is in its design phase.


\end{document}